\documentclass[altaffilletter,aps,nofootinbib,twocolumn,prd,eqsecnum,preprintnumbers,superscriptaddress,10pt,floatfix]{revtex4-2}
\pdfoutput=1
\usepackage{graphicx}
\usepackage{comment}
\usepackage{amsmath}
\usepackage{subfigure}
\usepackage{amssymb}
\usepackage{amsfonts}
\usepackage{mathtools}
\usepackage{amssymb}
\usepackage{enumerate}
\usepackage{xcolor}
\usepackage{bm}
\usepackage{mathrsfs}
\usepackage{epstopdf}
\usepackage{url}
\usepackage{footnote}
\usepackage{textcomp}
\usepackage{dsfont}
\usepackage{ulem}
\usepackage{hyperref}
\usepackage{enumerate}   
\usepackage{appendix}
\usepackage{textcomp}
\usepackage{tipa}

\makeatletter
\newcommand*{\rom}[1]{\expandafter\@slowromancap\romannumeral #1@}
\makeatother

\begin{document}

\title{Gravitational-wave Echoes Beyond the Hubble Time from Quantum Black Holes}

\author{Che-Yu Chen}
\email{b97202056@gmail.com}
\affiliation{RIKEN iTHEMS, Wako, Saitama 351-0198, Japan}

\author{Naritaka Oshita}
\email{oshita@phys.kindai.ac.jp}
\affiliation{Department of Physics, Kindai University, Osaka 577-8502, Japan}
\affiliation{RIKEN iTHEMS, Wako, Saitama 351-0198, Japan}

\author{Yuki Yokokura}
\email{yokokura.yuki@kochi-tech.ac.jp}
\affiliation{School of Engineering Science, Kochi University of Technology, 185, Miyanoguchi, Tosayamada, Kami, Kochi 782-5232, Japan}
\affiliation{RIKEN iTHEMS, Wako, Saitama 351-0198, Japan}

\begin{abstract}
We investigate gravitational-wave echoes in a horizonless compact object model, a non-perturbative solution of the semi-classical Einstein equation that reproduces the entropy-area law from the bulk structure. By employing a model of anisotropic fluids in classical general relativity (GR), we show that odd-parity gravitational waves are not reflected at the object's surface but are trapped inside due to the exponentially strong redshift, and that the echo-delay time extremely exceeds the age of the Universe. Consequently, the echo signals are effectively unobservable, which is consistent with current null results from echo searches. We finally discuss how this quantum black hole and the conventional classical black hole could be distinguished observationally. 
\end{abstract}

\maketitle

\section{Introduction}\label{sec.intro}

Black holes are mysterious astrophysical objects that realize strong gravitational fields. 
They emit gravitational waves that are detectable even at cosmological distances, and their extreme spacetime curvature is believed to lead to various quantum effects such as particle creation. Therefore, black holes are regarded as excellent laboratories for probing physics in extreme gravitational environments and for testing the quantum nature of gravity. However, the true nature of black holes remains unclear. Theoretically, three problems—singularities, the entropy-area law, and the information problem—must be resolved. Observationally, the existence of the event horizon has not yet been fully confirmed, leaving various possible alternative scenarios \cite{Cardoso:2019rvt,Bambi:2025wjx,Solodukhin:2025opw,Battista:2026nsx}. A theoretically and observationally consistent understanding of black holes is desired.

Currently, a variety of quantum-inspired models that deviate from classical black holes have been proposed \cite{Platania:2023srt,DelPiano:2023fiw,Zhang:2023yps,Battista:2023iyu,Calmet:2019eof,Chen:2023wdg,Capozziello:2024ucm,Belfaqih:2024vfk,Brahma:2020eos,Xiang:2013sza}. 
In recent years, observations of gravitational waves from black hole mergers \cite{LIGOScientific:2016aoc,LIGOScientific:2026sit} and images of black hole shadows \cite{EventHorizonTelescope:2019dse,EventHorizonTelescope:2022wkp} have opened new avenues for distinguishing between these models. 
In particular, gravitational-wave echoes have attracted significant attention as potential signals carrying information from the near-horizon region, arriving with a delay after the primary ringdown waveform \cite{Cardoso:2016rao,Cardoso:2016oxy,Cardoso:2017cqb}.

If gravitational waves interact with the surface of a quantum black hole that effectively possesses a nonzero reflectivity, one expects the appearance of echoes following the ringdown phase with separation time typically of the order of milliseconds for a solar-mass object. These echo waveforms can be constructed either from specific metric models, such as wormholes \cite{Cardoso:2016rao} or fuzzballs \cite{Ikeda:2021uvc} (see also Ref.~\cite{Cardoso:2016oxy}), or more phenomenologically by directly assuming a reflection coefficient \cite{Maggio:2017ivp,Hod:2017cga,Nakano:2017fvh,Mark:2017dnq} or modified dispersion relations in the ultraviolet regime \cite{Oshita:2018fqu} (for review papers, see e.g., Refs.~\cite{Cardoso:2019rvt,Abedi:2020ujo,Berti:2025hly}).

A number of black hole models proposed in previous studies---particularly those in which beyond-GR effects become significant near the horizon—have been argued to produce such echoes. On the other hand, recent observations by the LIGO–Virgo–KAGRA collaboration have achieved increasingly precise measurements of ringdown signals, making the empirical investigation of gravitational-wave echoes progressively more feasible, although no significant evidence for gravitational-wave echoes has been found yet \cite{LIGOScientific:2026wpt}.

In this context, the present work addresses the following question:
If gravitational-wave echoes are not observed, does this immediately imply that the observed object is (or very similar to) the classical black hole? The answer is no, provided that there exists even a single model that has a very different interior structure and does not \textit{effectively} emit echoes. 

In this paper, we prove this possibility in a quantum black hole model. Specifically, we consider a model proposed in Refs.~\cite{Kawai:2014afa,Kawai:2015uya,Kawai:2017txu,Kawai:2020rmt,Kawai:2021qdk,Yokokura:2022phq,Yokokura:2023wxp}, which we refer to as the ``KY model" in this paper. It is a non-perturbative solution (for the Planck constant) to the semi-classical Einstein equation; it describes a highly dense compact object consisting of excited quanta,
with a surface outside the Schwarzschild radius by the Planck length, rather than a horizon. See sec.~\ref{sec.meco} for the metric of the KY model.

This model has various interesting properties. First, the interior metric connects to the exterior Schwarzschild metric at the surface as smoothly as possible so that it is consistent with thermodynamics and Israel's junction condition \cite{Yokokura:2022phq}. Second, the curvature, which is close to the Planck scale inside the surface, induces quantum fluctuations, generating a large tangential pressure against the strong self-gravity, and the classical singularity disappears \cite{Kawai:2020rmt}. Third, considering its formation process, Hawking-like radiation occurs in the interior, and the object evaporates (except for the final stage of evaporation) \cite{Kawai:2013mda, Kawai:2020rmt}. Fourth, summing up the entropy density of the self-gravitating matter fields in the bulk interior reproduces the Bekenstein-Hawking formula exactly \cite{Yokokura:2022phq,Yokokura:2023wxp}. Fifth, although it has no horizon, the model has an exponentially large redshift, and the strong time-delay effect acting on the light rays propagating inside effectively darkens the image significantly; the image of the KY model is almost identical to a classical black hole image \cite{Chen:2024ibc}. Based on the above points, the KY model can be considered as a candidate for a quantum black hole. 

The main aim of the present paper is to investigate the dynamics of waves propagating in the KY model—a first-principles-based model—in order to further verify its consistency with observational data. Originally, the dynamics of the system is governed by the semi-classical Einstein equation, and its time-dependent analysis involves perturbations in both the metric and quantum states. As a first step to such a complete yet complicated approach, we here consider an effective analysis. In particular, for gravitational waves, we focus only on the axial (or parity-odd) sector of the metric perturbations by assuming that the KY geometry is described by GR minimally coupled to frictionless anisotropic fluids with negligible viscosity. This fluid assumption is consistent with a recent finding \cite{Yokokura_viscosity} that the ratio of shear viscosity $\mu$ and entropy density $\tilde{s}$ is strongly suppressed down to the fundamental quantum limit \cite{Kovtun:2004de}, $\mu/\tilde{s}\sim \hbar$, making macroscopic viscous dissipation entirely negligible.

In this setup, the wave equation and the associated effective potential are completely determined by the KY metric, which can consistently cover the entire domain ranging from the center to infinity, without introducing any extra phenomenological reflectivity such as reflection coefficients at the surface. Our results show that, first, the extremely compact configuration of the KY model ensures that the gravitational effects outside the object are basically the same as the classical case, meaning that the typical ringdown signals and the associated quasi-normal modes (QNMs) remain very close to those in the Schwarzschild case. Second, because the interior and exterior regions are connected almost smoothly, gravitational waves can easily enter the surface and propagate inside with nearly no reflection at the surface. The waves are reflected by the centrifugal potential barrier around the center, propagate outward, and escape from the surface to the exterior. A portion of these waves would be observed as echo signals. However, due to the exponentially strong redshifts of the KY model, these signals will be delayed significantly, and the echo-separation timescale is far longer than the Hubble time; \textit{practically, the echoes cannot be detected.} Therefore, at the level of the current gravitational wave observations, the KY model remains consistent with the phenomenological behavior of classical black holes, even though the interior structure is drastically different. We therefore argue that, in terms of testing quantum black hole models that exhibit such extreme redshift effects, it is necessary to consider probes other than (axial-sector) gravitational-wave echoes.

This paper is organized as follows. In sec.~\ref{sec.meco}, we briefly review the KY model. In sec.~\ref{sec:waveeq}, we present the wave equation and the associated effective potential, from which the waveforms are numerically calculated and analyzed in sec.~\ref{sec:numer}. We finally conclude in sec.~\ref{sec:conclusion}.

\section{KY model}\label{sec.meco}

\begin{figure}[t]
  \centering
  \includegraphics[scale=0.1]{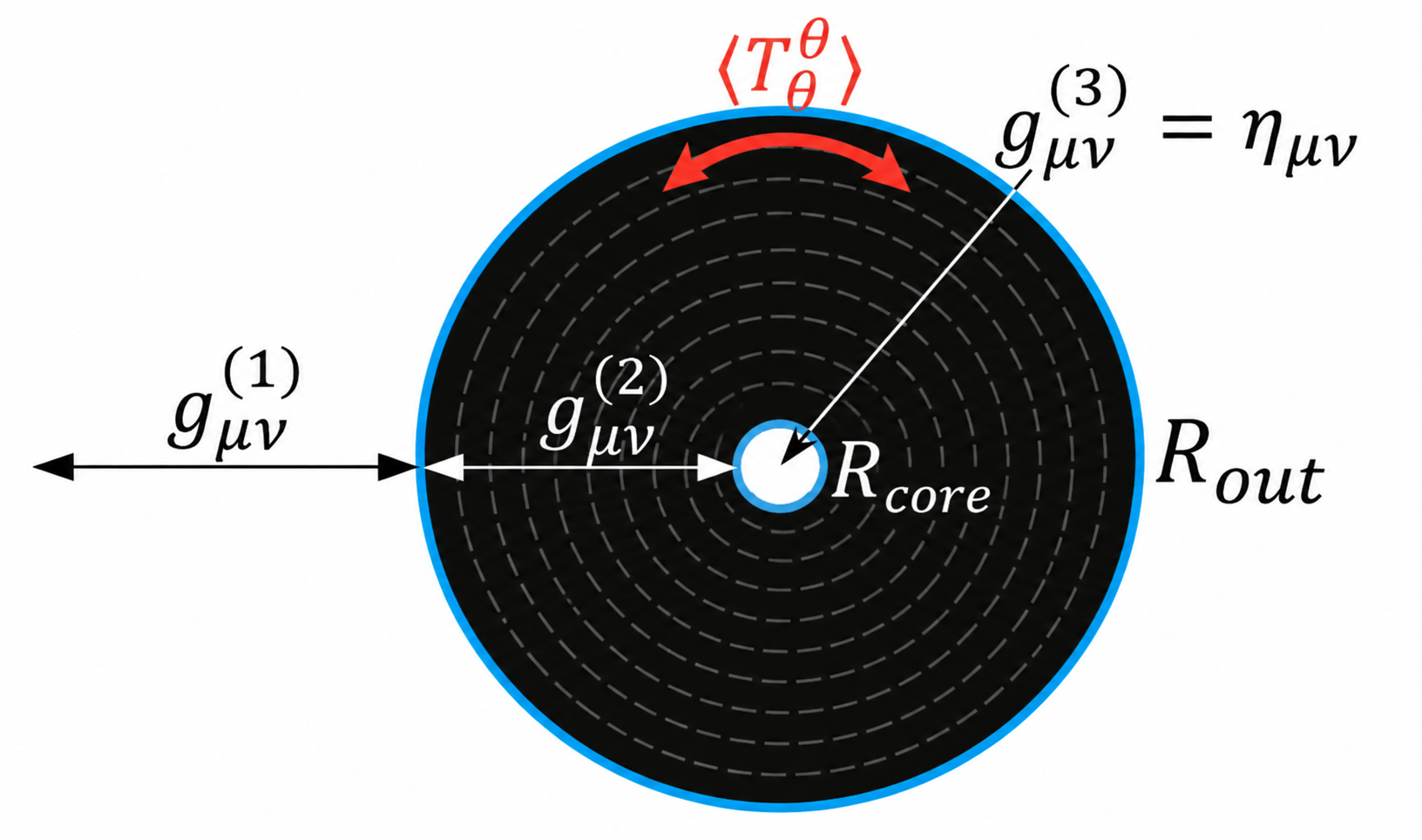}
\\
\caption{The configuration of the KY model. It contains three regions, and there is no event horizon. The geometry inside is supported by the large tangential quantum pressure $\langle T^\theta{}_\theta \rangle$.}
\label{fig:structure} 
\end{figure}

The geometry of the KY model consists of three parts: the exterior Schwarzschild part, the interior dense part, and the central core part. The first two are connected at the surface $r=R_{\rm out}$, and the latter two at $r=R_{\rm core}$. See Fig.~\ref{fig:structure} for the configuration of the object. 

More explicitly, the KY spacetime can be described by the following spherically symmetric and static metric \cite{Kawai:2013mda,Kawai:2014afa,Kawai:2015uya,Kawai:2017txu,Kawai:2020rmt,Kawai:2021qdk,Yokokura:2022phq,Yokokura:2023wxp}:
\begin{equation}
ds^2=-A_{i}(r_i)dt_{i}^2+B_{i}(r_i)dr_i^2+r_i^2d\Omega_2^2\,,\label{metric}
\end{equation}
where the subscripts $i=1$, $2$, and $3$ label the quantities in the exterior region, the dense configuration, and the central core, respectively.

The spacetime outside the surface located at $r=R_\textrm{out}$ is approximately described by the Schwarzschild metric:
\begin{equation}
A_1\left(r_1\right)=\frac{1}{B_1\left(r_1\right)}=1-\frac{a_0}{r_1}\,,\label{schwar}
\end{equation}
where $a_0$ is the Schwarzschild radius for the Arnowitt-Deser–Misner energy $a_0/2G$ with $G$ the gravitational constant.

The metric functions in the dense configuration for $R_{\rm core}\leq r \leq R_{\rm out}$ are expressed as
\begin{align}
A_2\left(r_2\right)&=\left(1-\frac{a_0}{R_\textrm{out}}\right)\frac{R_\textrm{out}^2}{r_2^2}\,\textrm{exp}\left({\frac{r_2^2-R_\textrm{out}^2}{2\sigma\eta}}\right)\,,\nonumber\\
B_2\left(r_2\right)&=\frac{r_2^2}{2\sigma}\,,
\label{dense_metric}
\end{align}
characterized by two parameters $(\sigma,\eta)$ and the surface radius $R_\textrm{out}$. Here, $\sigma$ corresponds to the intensity of Hawking radiation, and its order is $\sigma\sim n \ell_\textrm{Pl}^2$, where $\ell_\textrm{Pl}\equiv \sqrt{\hbar G}$ is the Planck length, $n$ is the number of degrees of freedom that contribute to the radiation and the entropy, and $n$ is assumed to be $\mathcal{O}(1)$ but large (say, $n\sim 1000$) such that fluctuations of geometry are suppressed \cite{Kawai:2013mda,Kawai:2014afa,Yokokura:2022phq}. The other parameter $\eta$ is a phenomenological parameter that characterizes interaction between quanta inside \cite{Kawai:2014afa}, whose value is restricted to $1\le\eta<2$ from causality and thermodynamics \cite{Kawai:2020rmt, Yokokura:2022phq}. More precisely, $\sigma$ and $\eta$ are determined by solving the semi-classical Einstein equation self-consistently, and they are related to each other as \cite{Kawai:2014afa,Kawai:2020rmt}
\begin{equation}\label{sigma_eta}
    \sigma=\frac{cn \ell_\textrm{Pl}^2}{\eta^2}\,,
\end{equation}
where $c$ is an $\mathcal{O}(1)$ numerical coefficient.
Note that the redshift in Eq.~\eqref{dense_metric} is exponentially strong for $r<R_{\rm out}-\sigma/R_{\rm out}$.

At this point, the relation between $R_\textrm{out}$ and $a_0$ remains undetermined. It can be fixed by Israel's junction condition or a thermodynamic condition, which leads to the continuity of the proper acceleration
\begin{equation}\label{acceleration}
\alpha(r)\equiv\frac{\partial_r\ln{\sqrt{-g_{tt}(r)}}}{\sqrt{g_{rr}(r)}}
\end{equation}
at the surface $R_\textrm{out}$: $\alpha_1(R_\textrm{out})=\alpha_2(R_\textrm{out})$ \cite{Yokokura:2022phq}.\footnote{The geometric definition of the proper acceleration is $\alpha\equiv |g_{\mu\nu}\alpha^\mu \alpha^\nu|^{1/2}$, where $\alpha^\mu=u^\nu \nabla_\nu u^\mu$ and $u^\mu \partial_\mu =\frac{1}{\sqrt{-g_{tt}}}\partial_t$. For a spherically symmetric and static metric, this agrees with \eqref{acceleration}, and it is the acceleration required to stay at a point $r$ against the self-gravity.} 
This gives the following algebraic relation 
\begin{equation}
a_0^2R_\textrm{out}\sigma\eta^2-2\left(R_\textrm{out}-a_0\right)\left(R_\textrm{out}^2-2\eta\sigma\right)^2=0\,,\label{algebraicRa0}
\end{equation}
from which one can obtain the analytic relation $a_0=a_0(R_\textrm{out})$ as
\begin{equation}
a_0=a_0(R_\textrm{out})=\frac{-W^2+\sqrt{W^4+2R_\textrm{out}^2W^2\sigma\eta^2}}{R_\textrm{out}\sigma\eta^2}\,,\label{Rexp}
\end{equation}
with $W\equiv R_\textrm{out}^2-2\eta\sigma$. For $\sigma\ll R_\textrm{out}^2$, one can expand Eq.~\eqref{Rexp} and obtain the approximated relation \cite{Yokokura:2022phq}:
\begin{equation}
R_\textrm{out}\approx a_0+\frac{\sigma\eta^2}{2a_0}\,.\label{Raapp}
\end{equation}
Noting again that $a_0$ is the Schwarzschild radius and that $\sigma \eta^2\sim n \ell_\textrm{Pl}^2$ from \eqref{sigma_eta}, 
the relation \eqref{Raapp} means that the KY model is very compact. This is a consequence of the large tangential pressure $\langle T^\theta{}_\theta \rangle$ non-perturbatively induced by 4D quantum fluctuations of matter fields \cite{Kawai:2020rmt}.

Finally, let us specify the central part. The dense metric \eqref{dense_metric} with the parameter \eqref{sigma_eta} is a non-perturbative solution to the semi-classical Einstein equation for a large value of $n$, and it is not applicable to a small central region of size $\sim \ell_\textrm{Pl}$, where the quantum-field-theory description breaks down \cite{Yokokura:2023wxp}. In this sense, the central small region cannot be controlled by the semi-classical dynamics.\footnote{Although the dense metric is not applicable to the region $0\leq r \lesssim \ell_\textrm{Pl}$, we can check that it can contain at most the Planck mass \cite{Yokokura:2022phq,Yokokura:2023wxp}, meaning that there is no solar-mass scale singularity. Recently, an improved model has been obtained in an effective Hamiltonian approach \cite{Livine:2025soz}; it resolves the singularity completely and matches the metric of the original model for $r \gg \ell_\textrm{Pl}$.}  As will be seen in the following sections, however, the main result -- gravitational-wave echoes from the KY model are effectively undetectable -- does not depend on the details of the central region (due to the large redshift in the interior dense part) as long as it is regular. 

We here assume, for simplicity, that the small central part is flat, and the inner surface is chosen as 
\cite{Kawai:2020rmt,Yokokura:2023wxp}
\begin{equation}
    R_\textrm{core}=\sqrt{2\sigma}\,.
\end{equation}
By connecting to the dense metric \eqref{dense_metric} at $r=R_\textrm{core}$, the flat metric for the region $0\leq r \leq R_{\rm core}$ is given by 
\begin{align}
A_3\left(r_3\right)&=\left(1-\frac{a_0}{R_\textrm{out}}\right)\frac{R_\textrm{out}^2}{2\sigma}\,\textrm{exp}\left({\frac{2\sigma-R_\textrm{out}^2}{2\sigma\eta}}\right)=\textrm{const.}\,,\nonumber\\
B_3\left(r_3\right)&=1\,.\label{metricint3}
\end{align}
Thus, these three metrics, i.e., Eqs.~\eqref{schwar}, \eqref{dense_metric}, and \eqref{metricint3}, together describe the whole of spacetime. 

Before concluding this section, we introduce the following dimensionless quantity for later analysis:
\begin{equation}\label{k_def}
k\equiv \frac{\sigma\eta^2}{a_0^2}\sim\frac{n\ell_\textrm{Pl}^2}{a_0^2}\,,
\end{equation}
where we applied the relation \eqref{sigma_eta}. 
Noting the second term in the relation \eqref{Raapp}, the parameter $k$ can roughly quantify the deviation in size from the classical case due to quantum effects, i.e., $R_\textrm{out}/a_0\approx 1+k/2$. For a solar-mass size and $n\sim1000$, we have $k\sim10^{-74}$, which is very small. Keeping this in mind, in the following analysis, we will use a fairly large value for $k$ to make the results of our numerical calculations easier to visualize.

\section{Wave equations}\label{sec:waveeq}

As introduced in the previous section, the KY model is characterized by exponentially strong redshifts inside the dense configuration. In this section and in the next section, we will show that the wave dynamics, e.g., scalar fields, electromagnetic waves, and gravitational waves propagating through the interior of the KY model, gives rise to echoes after the ringdown signals. Since the ringdown is essentially determined by the wave-scattering process around the unstable photon sphere (light ring), which remains intact in this model, the ringdown signal of the KY model is almost identical to that of the classical case. However, since this model lacks an event horizon, additional processes occur subsequently. A portion of the waves entering the object is reflected at the core, propagates outward, and is reflected again by the light ring. By repeating this process, a series of echoes can be generated. Here, the separation timescale between echoes depends on the redshift experienced during their propagation inside the object. The main goal in this paper is to show how the strong gravitational redshift inside the KY model would have some implications on the echo-separation timescale.

We will focus on the wave dynamics of massless scalar fields, electromagnetic fields, and gravitational waves. For the scalar field and electromagnetic fields, we treat them as waves propagating in a fixed KY spacetime and neglect the backreaction of waves acting on the spacetime geometry. For gravitational waves, we only focus on the parity-odd sector, as a first step, by assuming that the linear perturbation dynamics of the KY spacetime can be modeled by a spherically symmetric configuration in GR minimally coupled to frictionless anisotropic fluids (see Appendix \ref{appendixpotential}), since it can capture the main feature of the KY model, i.e., the extreme delay of the echo-separation timescale due to strong redshifts. We expect that the result should be robust even if a self-consistent gravitational perturbation based on the semi-classical Einstein equation is carried out, which is beyond the scope of the present study.

In this setup, the master equations for the wave dynamics propagating in the KY model can be expressed as
\begin{equation}
\frac{\partial^2\Psi}{\partial t^2}-\frac{\partial^2\Psi}{\partial x^2}+V(r)\Psi=0\,,\label{waveeq}
\end{equation}
where the tortoise radius $x$ is defined as
\begin{equation}
dx=\sqrt{\frac{B(r)}{A(r)}}dr\,,\label{tortoise}
\end{equation}
and we dropped the subscript $i$ for simplicity. The wavefunction $\Psi$ stands for the proxy of the scalar waves, electromagnetic waves, or gravitational perturbations under consideration. The potential $V(r)$ is given by \cite{Chen:2019iuo,Bouhmadi-Lopez:2020oia}
\begin{widetext}
\begin{equation}
V(r)=\frac{A(r)s(s-1)}{r^2B(r)}+\frac{A(r)\left[\ell(\ell+1)-s(s-1)\right]}{r^2}+\frac{(s-1)\left[B'(r)A(r)-B(r)A'(r)\right]}{2rB(r)^2}\,,\label{veff}
\end{equation}
\end{widetext}
with the multipole number $\ell$ and the spin of the field $s$. Note that the wavefunction $\Psi$ should be labeled by $\ell$, while we have omitted it for simplicity. The prime denotes the derivative with respect to $r$. For scalar fields, electromagnetic fields, and axial gravitational perturbations, we have $s=0$, $s=1$, and $s=2$, respectively. The detailed derivation of the master equation for $s=2$ is shown in Appendix \ref{appendixpotential}. Note that this potential in the eikonal limit, i.e., $\ell\gg1$, recovers the one for the photon geodesic equations multiplied by $\ell^2$ \cite{Chen:2024ibc}.
 
To visualize the potential as a function of $x$, we first integrate Eq.~\eqref{tortoise} in the exterior region to get $x(r)$ in the region $r\ge R_\textrm{out}$. The integration constant is set such that $x=2a_0$ when $r=2a_0$. The potential at the exterior is obtained using Eq.~\eqref{veff} with $r$ expressed by $x$. Then, we numerically integrate Eq.~\eqref{tortoise} in the dense region, i.e., Eq.~\eqref{dense_metric}, fixing the integration constant by matching the interior tortoise coordinate with the exterior one at $r=R_\textrm{out}$; the potential in the dense region is given by Eq.~\eqref{veff} with $r$ expressed as a function of $x$. The potential inside the core region is obtained in a similar manner and can be connected to the dense region at $r=R_\textrm{core}$.

\begin{figure*}
  \centering
  \subfigure[]
{ \includegraphics[scale=0.35]{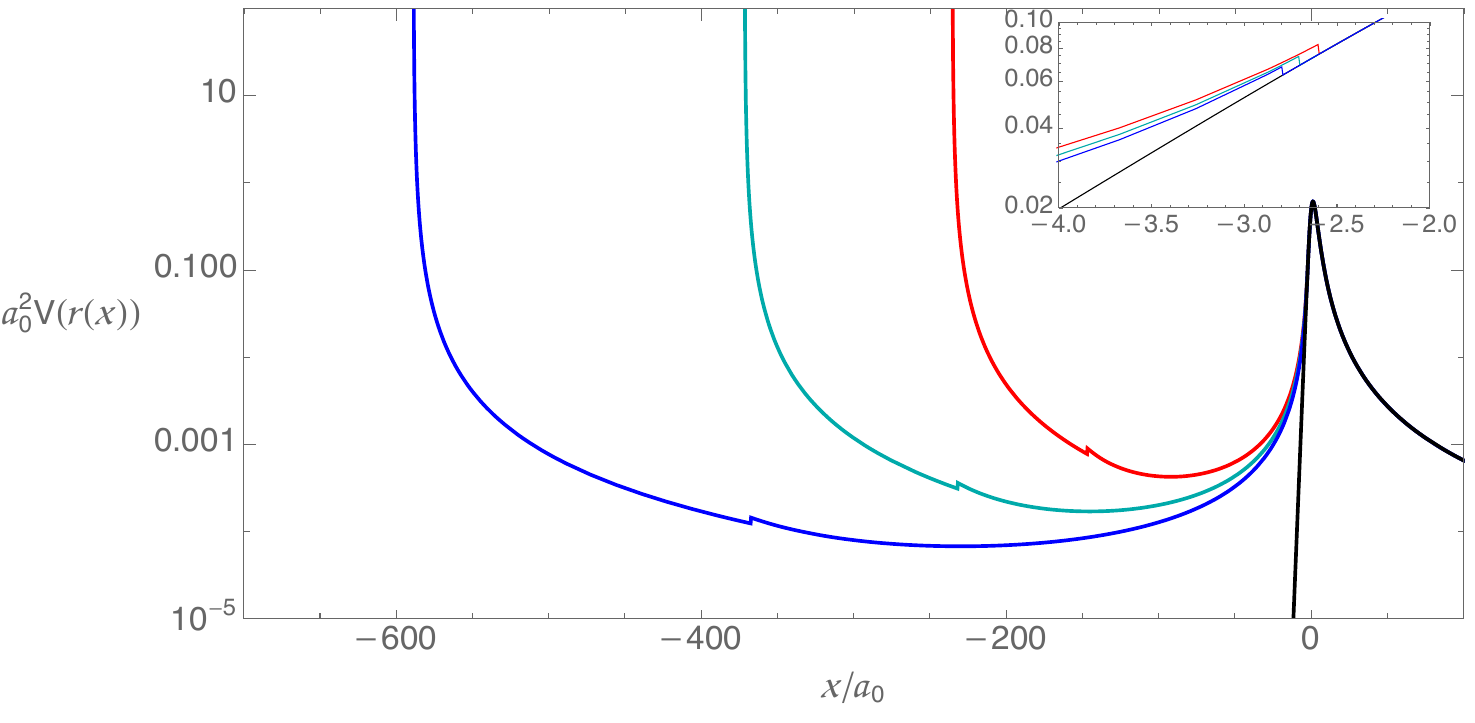}}
\subfigure[]
  {\includegraphics[scale=0.35]{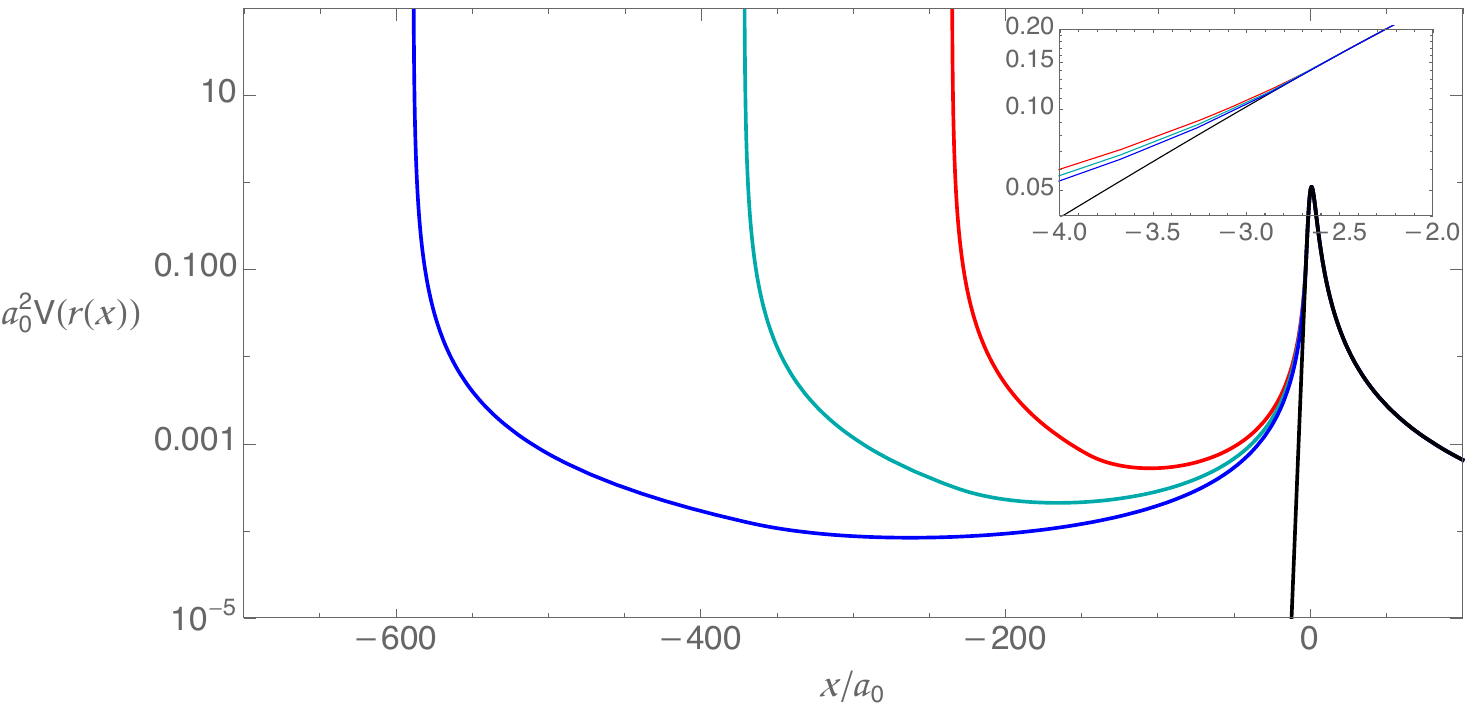}}
\\
\caption{(a) The effective potentials for gravitational perturbations $(s=2)$ and (b) for electromagnetic waves $(s=1)$. We fix $\ell=2$ and $\eta=1$. The quantum parameter $k$ is chosen to be $k=1/21$ (red), $k=1/23$ (green), and $k=1/25$ (blue). The black curve shows the potential for the Schwarzschild black hole.}
\label{fig:potentials} 
\end{figure*} 

In Fig.~\ref{fig:potentials}, we show the potential $V(r(x))$ for the gravitational perturbations ($s=2$) and the electromagnetic fields $(s=1)$ in the left and the right panels, respectively. We choose $\ell=2$ and $\eta=1$, then vary the quantum parameter $k$ as $k=1/21$ (red), $k=1/23$ (green), and $k=1/25$ (blue). The black curve represents the standard Schwarzschild potential; it has a peak near the light ring $r/a_0=3/2$, decreases rapidly inward from there, and vanishes at $r=a_0$. In the potential of the KY model, on the other hand, $R_\textrm{out}$ is located at the point where the colored curves and the black curve start to differ (see the insets). The core surface $R_\textrm{core}$ corresponds to the point for the small jump around the left of the local minimum in the left panel. The potential value at the local minimum between $R_{\rm core}$ and $R_{\rm out}$ is exponentially small when choosing a small value of $k$ because its value is dominantly suppressed by the exponential factor in $A(r)$. In addition, inside the core region, there is an infinitely high potential barrier due to centrifugal forces. The separation between this infinitely high potential barrier and the exterior peak, in units of $x/a_0$, significantly increases when decreasing $k$.

We also notice that the effective potential may have tiny discontinuities both at the KY surface $R_\textrm{out}$ and the core surface $R_\textrm{core}$. These come from the discontinuities of $A'(r)$ and $B'(r)$ in the last term on the right-hand side of Eq.~\eqref{veff}, and only appear when $s\ne1$. We evaluate the discontinuities of the effective potential, i.e., $\Delta V(R)\equiv \lim_{r\to R^+}V(r)-\lim_{r\to R^-}V(r)$, where $R=R_{\rm out},\,R_{\rm core}$. 
In the limit of small $k$, these discontinuities read
\begin{align}
a_0^2\Delta V(R_\textrm{out})&=-\frac{(s-1)(\eta-1)k}{2\eta}\nonumber\\
&+\frac{(s-1)(\eta-2)[7\eta+s(2+\eta)]k^2}{4\eta^2}+\mathcal{O}(k^3)\,,\\
a_0^2\Delta V(R_\textrm{core})&\approx \frac{(s-1)(2\eta-1)\eta^3}{8k}e^{-\eta/2k}\,,
\end{align}
which are suppressed linearly and exponentially in $k\rightarrow0$, respectively. The discontinuity at $r=R_{\rm core}$ 
is visible in Fig.~\ref{fig:potentials}(a), i.e., near the potential minimum because of the logarithmic scale in the vertical axis. On the other hand, the discontinuity at $r=R_{\rm out}$ 
is visible only through the small inset in Fig.~\ref{fig:potentials}(a). Generically, the existence of discontinuities of the effective potential may induce extra echo signals due to fractional reflection at the discontinuities. However, in the KY model, the discontinuities both at $r=R_{\rm core}$ and $R_\textrm{out}$ are suppressed by a small $k$. No significant echoes associated with these discontinuities would appear unless extra reflection mechanisms are imposed there.

Finally, we emphasize again that the discontinuities do not appear when $s=1$ because the effective potential is simply given by $\propto A(r)/r^2$ in this case, and the metric function $A(r)$ is continuous. Therefore, these tiny discontinuities do not appear in the electromagnetic case (Fig.~\ref{fig:potentials}(b)).

\section{Numerical results}\label{sec:numer}

In sec.~\ref{sec:waveeq}, we have presented the master equation for the wave dynamics \eqref{waveeq} and the effective potential \eqref{veff}. In this section, we will analyze them numerically to demonstrate the echoes. In particular, we will solve the equation \eqref{waveeq} from both the frequency domain and the time domain. The results obtained by these two approaches are consistent with each other, which ensures the validity of the results.

\subsection{Frequency-domain analysis}
In the first approach, we perform a numerical computation to see how a scattered waveform excites QNMs and echoes. Our computation is based on the Green's function in the frequency space.

We utilize the ansatz
\begin{equation}
\Psi = \int d \omega e^{-i \omega t} \psi_{\omega} (x)\,.
\label{sph_decomposition}
\end{equation}
Given the initial data $\Psi (t=0, x)$ and $ \partial \Psi (t,x) / \partial t|_{t = 0}$, Eq.~\eqref{waveeq} reduces to the Schr\"{o}dinger-type equation \cite{Regge:1957td,Zerilli:1970se}:
\begin{equation}
\left[ \frac{d^2}{d x^2} +\omega^2 -V(r(x)) \right] \psi_{\omega} (x) = I(\omega, x)\,,
\end{equation}
where the source term includes the initial data
\begin{equation}
    I (\omega, x) = i \omega \Psi (t=0, x) -\partial \Psi (t,x) / \partial t|_{t = 0}\,.
    \label{source_initial}
\end{equation}
One can solve this equation by utilizing the Green's function. It is obtained by imposing the following boundary conditions:\footnote{In practice, one has to impose the regularity condition at a finite value of the tortoise coordinate, $x=x_{\rm c}$, when performing numerical computations.
We impose the boundary condition of $\psi_{\omega}(x_{\rm c}) = C \ (>0)$ and $\partial_x \psi_{\omega}(x_{\rm c}) = 0$ with $-x_c \gg 1$, where $C$ is a constant.
Note that the full solution is invariant under an overall rescaling of the homogeneous solution.
}
\begin{align}
\psi_{\omega} = \partial_x \psi_{\omega} = 0 \  &\text{for} \ r =0\,,  \label{inbc}\\
\psi_{\omega} = e^{i \omega x} \ &\text{for} \ r \to \infty\,.
\label{outbc}
\end{align}
One can construct the Green's function with the two independent homogeneous solutions $\psi_{\omega}^{\rm (in)} (x)$ and $\psi_{\omega}^{\rm (out)} (x)$, which satisfy the boundary conditions, $\psi_{\omega}^{\rm (in)} (x) = 0$ at $r=0$ and $\psi_{\omega}^{\rm (out)} (x) = e^{i \omega x}$ at $r\to \infty$, respectively. The generic behavior of $\psi_{\omega}^{\rm (in)} (x)$ for $r\to \infty$ is the mixture of ingoing and outgoing plane waves:
\begin{equation}
\psi_{ \omega}^{\rm (in)} (x)=A_{\rm in} (\omega) e^{-i \omega x} + A_{\rm out} (\omega) e^{i \omega x} \quad \text{for} \ r \to \infty\,.\label{inbranch}
\end{equation}
Then, the Green's function can be written as 
\begin{equation}
G(x, x') = \frac{1}{W(\omega)} 
\begin{cases}
    \psi_{\omega}^{\rm (in)} (x) \psi_{\omega}^{\rm (out)} (x') \ \text{for} \ x \leq x'\,,\\
    \psi_{\omega}^{\rm (in)} (x') \psi_{\omega}^{\rm (out)} (x) \ \text{for} \ x \geq x'\,,
\end{cases}\label{greenfunction}
\end{equation}
which ensures the continuity of $G(x,x')$ at $x=x'$. The prefactor $W(\omega)$ can be determined by the condition $\partial G(x,x')/\partial x\Big|^{x\rightarrow x'^+}_{x\rightarrow x'^-}=1$, and one gets
\begin{equation}
W(\omega) \coloneqq \frac{\partial \psi_{\omega}^{\rm (out)}}{\partial x} \psi_{\omega}^{\rm (in)} - \frac{\partial \psi_{\omega}^{\rm (in)}}{\partial x} \psi_{\omega}^{\rm (out)}\,.
\end{equation}
Noting that $dW/dx=0$, one can easily find $W(\omega) = 2i \omega A_{\rm in}(\omega)$. 

When the observer is located at a far region, i.e., $x \to \infty$, the time domain waveform is obtained via $\psi_\omega(x)=\int dx'G(x,x')I(\omega,x')$, by applying the one for $x\ge x'$ in Eq.~\eqref{greenfunction}, as
\begin{equation}
\psi_{\omega} (x) = \frac{\psi_{\omega}^{\rm (out)} (x)}{2i \omega A_{\rm in} (\omega)} \int dx' \psi_{\omega}^{\rm (in)}(x') I(\omega, x')\,.
\end{equation}
If the initial data is localized at $x' \sim x_0$ with the hierarchy $2M \ll x_0 \ll x$ such that the plane-wave solutions are still valid around $x_0$, it is approximated as
\begin{equation}
\psi_{\omega} (x) = \frac{e^{i \omega x}}{2i \omega} \int dx' \left( e^{-i \omega x'} + \frac{A_{\rm out}}{A_{\rm in}} e^{i \omega x'} \right) I(\omega, x')\,.
\label{approx_spectra}
\end{equation}
The first term in Eq.~\eqref{approx_spectra} is relevant to the prompt response, i.e., a signal directly reaches the observer from the initial wavepacket. The second term is associated with the signal scattered at the light ring. Indeed, it has poles at $\omega = \omega_{\rm q}$ in the complex frequency plane, which are nothing but QNM frequencies, i.e., $A_{\rm in} (\omega_{\rm q}) = 0$. As we are interested in the excitation of QNMs and echoes, which are relevant to the second term in Eq.~\eqref{approx_spectra}, we will compute only the waveform associated with the second term in Eq.~\eqref{approx_spectra}. Its spectrum is denoted by $\psi_{\omega}^{\rm RD} (x)$:
\begin{equation}
\psi_{\omega}^{\rm RD} (x) = e^{i \omega x} \frac{A_{\rm out}}{2i\omega A_{\rm in}} \int dx' e^{i \omega x'} I (\omega, x')\,.
\end{equation}
The source term $I(\omega,x)$ can be obtained from the initial data [see Eq.~\eqref{source_initial}].
We here set the initial data as
\begin{align}
\Psi (t=0, x) &= 0\,,\\
\partial \Psi (t,x) / \partial t|_{t = 0} &= \exp \left[ - \frac{(x-x_0)^2}{\delta^2} \right]\,.
\end{align}

\begin{figure}
  \centering
\includegraphics[width=0.95\linewidth]{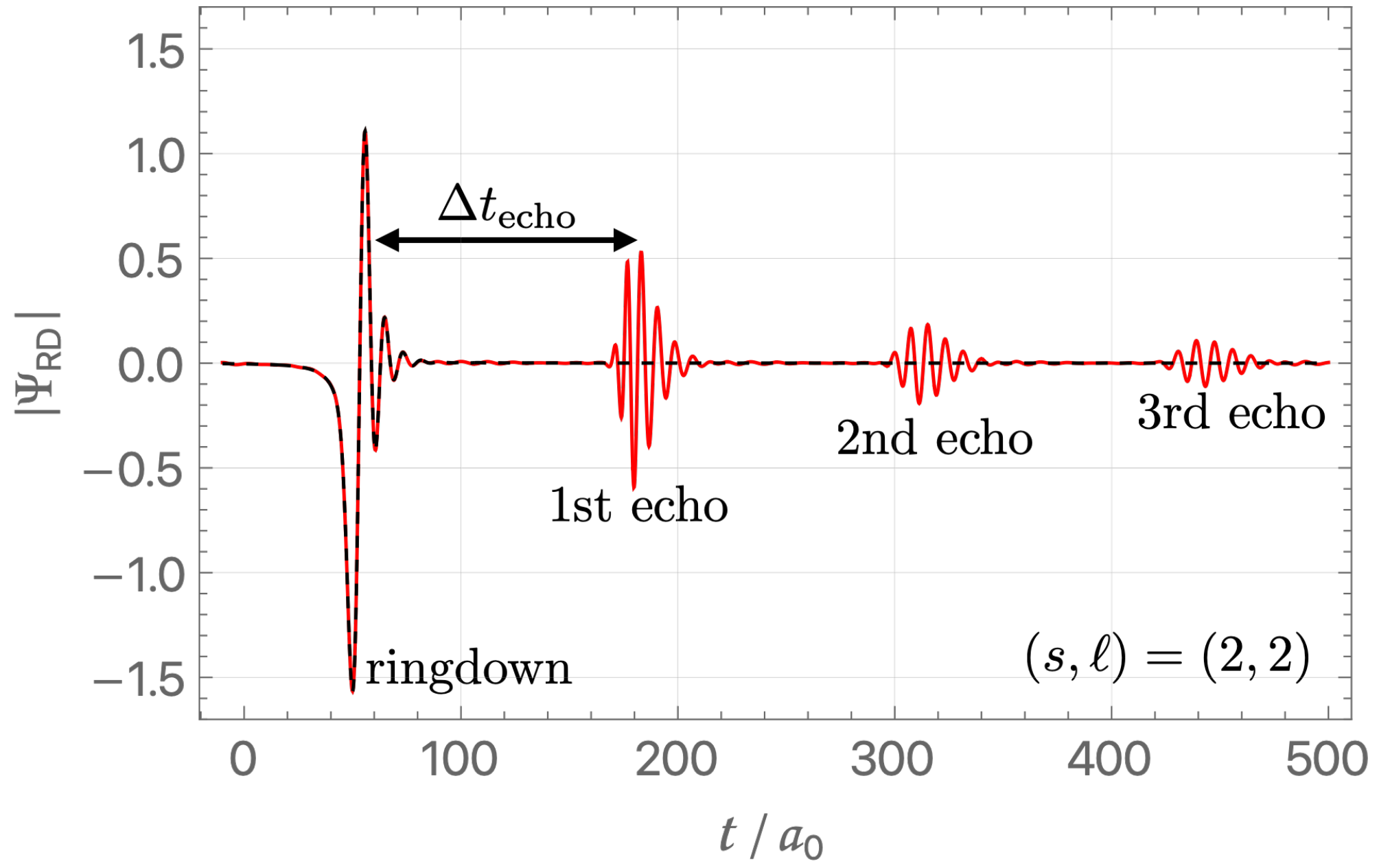}
\\
\caption{The waveforms for gravitational perturbations ($s=2$) based on the frequency-domain analysis \eqref{Psi_RD}. 
We set $\ell=2$ and $\eta=1$. 
The quantum parameter $k$ is chosen to be $k=1/15$ (red). 
The black dashed line shows the waveforms for the Schwarzschild black hole.}
\label{fig:echo_freq_s2} 
\end{figure}

We thus obtain the ringdown waveform by performing the inverse Laplace transform of $\psi_{\omega}^{\rm RD}$
\begin{equation}\label{Psi_RD}
    \Psi = \int_{- \infty}^{\infty} d \omega e^{-i \omega t} \psi_{\omega}^{\rm RD}(x)\,. 
\end{equation}
The result of ringdown and echoes for the gravitational perturbations ($s=2$) is shown in Fig.~\ref{fig:echo_freq_s2}.
We here set the model parameters as $k = 1/15$ and $\eta = 1$. The echo-separation time is around $\Delta t_{\rm echo} \simeq 2 \Delta x_{\rm sep} \simeq 120a_0$, where $\Delta x_{\rm sep}$ stands for the separation in $x$ between the infinitely high potential barrier and the exterior peak in the effective potential. Each echo appears after the waves travel back and forth between the potential barrier and the exterior peak, giving rise to the factor two in $\Delta t_{\rm echo}\simeq 2 \Delta x_{\rm sep}$.

Note that although we set $k=1/15$ here, a more realistic value of $k$ should be much smaller (see the definition of $k$ in Eq.~\eqref{k_def}), which will give rise to a much longer echo-separation timescale (see Eq.~\eqref{testimate}).

\subsection{Time-domain analysis}

The waveforms obtained through the frequency-domain analysis in the previous subsection clearly show the existence of echoes. In this subsection, we will solve the master equations \eqref{waveeq} purely in the time domain. In particular, we will show that the results obtained by the time-domain analysis agree with those obtained from the frequency-domain analysis, confirming the consistency of the two methods. In addition, by varying the values of the quantum parameter $k$, we will demonstrate how the waveform echoes are delayed when decreasing the values of $k$, i.e., when the redshifts become strong inside the KY model.   

\begin{figure*}
  \centering
  \subfigure[]
{ \includegraphics[scale=0.45]{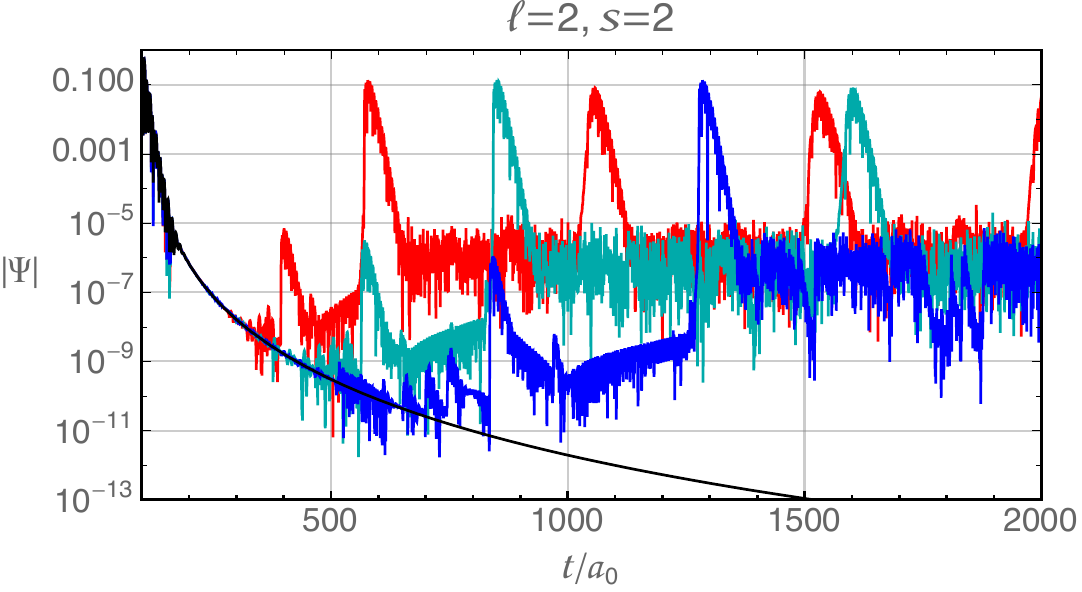}}
\subfigure[]
  {\includegraphics[scale=0.45]{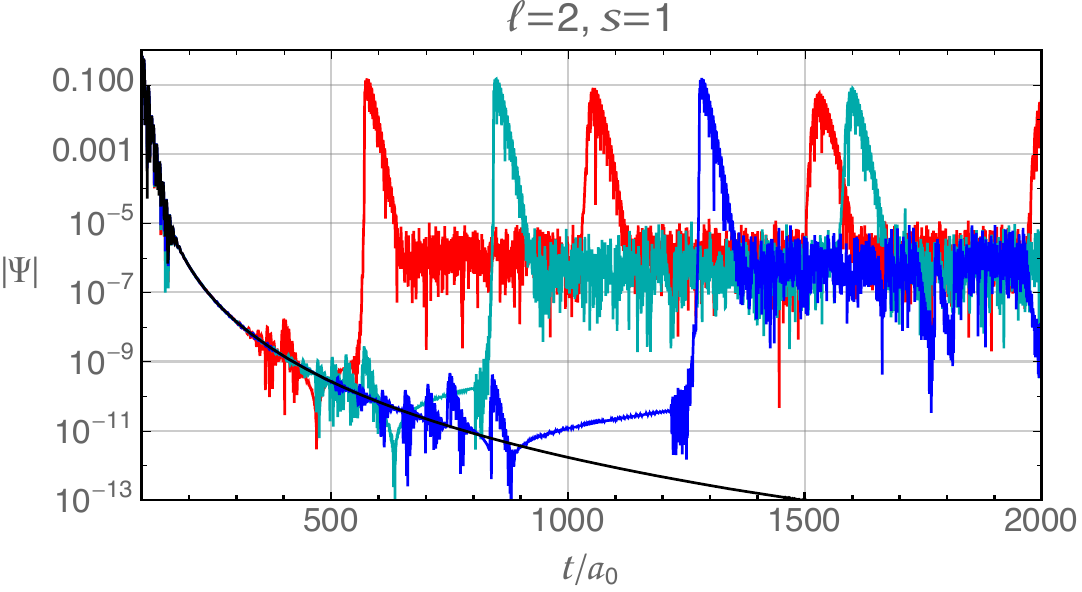}}
\\
\caption{(a) The waveforms for gravitational perturbations ($s=2$) and (b) for electromagnetic fields ($s=1$). We set $\ell=2$ and $\eta=1$. The quantum parameter $k$ is chosen to be $k=1/21$ (red), $k=1/23$ (green), and $k=1/25$ (blue). The black curve shows the ringdown waveforms for classical Schwarzschild black holes.}
\label{fig:waveform} 
\end{figure*}

\begin{figure*}
  \centering
  \subfigure[]
{ \includegraphics[scale=0.45]{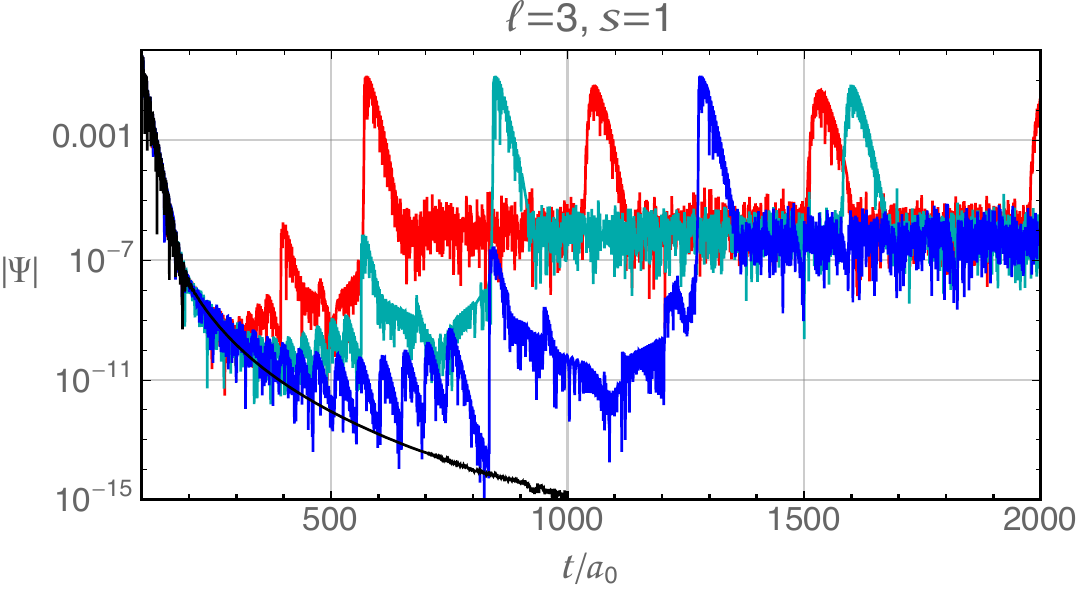}}
\subfigure[]
  {\includegraphics[scale=0.45]{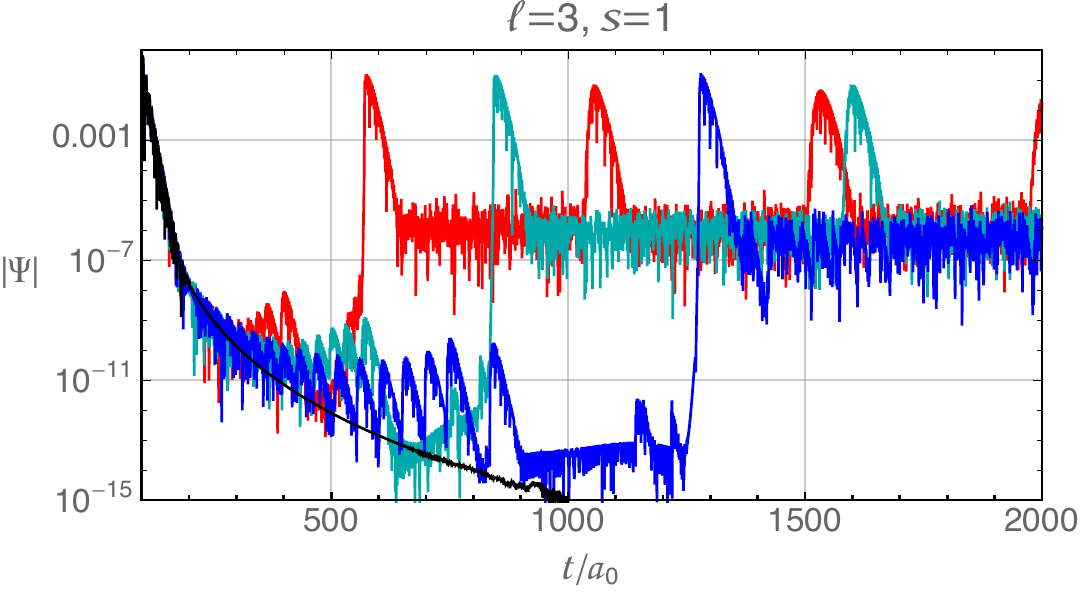}}
\\
\caption{The same setting as in Fig.~\ref{fig:waveform}, but with $\ell=3$.}
\label{fig:waveforml3} 
\end{figure*}

To numerically integrate the wave equation \eqref{waveeq} in the time domain, we first introduce the light cone coordinates
\begin{equation}
u=t-x\,,\qquad v=t+x\,.
\end{equation}
The wave equation \eqref{waveeq} can be rewritten as
\begin{equation}
4\frac{\partial^2\Psi}{\partial u\partial v}+V(r(u,v))\Psi=0\,.
\end{equation}
The boundary conditions for $\Psi(u,v)$ are given by
\begin{align}
\Psi(u,v=0)&=\exp \left[ - \frac{(-v_0)^2+(u-u_0)^2}{\Sigma^2} \right]\,,\nonumber\\
\Psi(u=0,v)&=\exp \left[ - \frac{(v-v_0)^2+(-u_0)^2}{\Sigma^2} \right]\,,
\end{align}
where $u_0$ and $v_0$ represent the location of the Gaussian peak on the $u$-$v$ plane.{\footnote{In this time-domain analysis, we fix $u_0=0$, $v_0=3a_0$, and $\Sigma=10 a_0$.}} Then, we numerically solve the wave equation in the $u$-$v$ plane. We assume that the waveform is read off by an observer located at $x=100a_0$. The results of the waveforms for $\ell=2$ and $\ell=3$ are shown in Figs.~\ref{fig:waveform} and \ref{fig:waveforml3}, respectively.

\begin{figure}[t]
  \centering
  \includegraphics[scale=0.48]{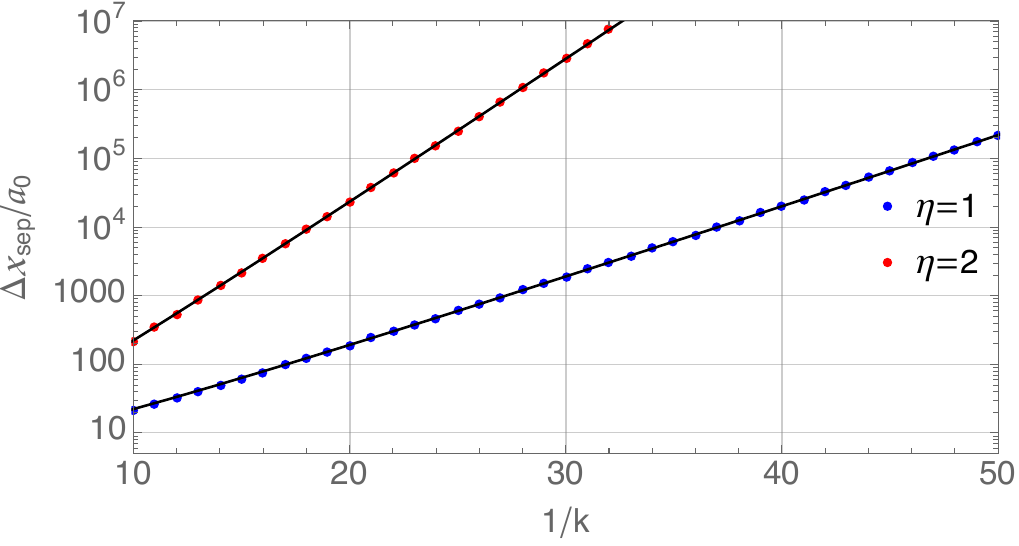}
\\
\caption{The peak separation $\Delta x_{\rm sep}$ as a function of $1/k$. The points are computed numerically from the potential \eqref{veff} with $\eta=1$ (blue) and $\eta=2$ (red). The black curves show how the numerical results agree with the approximated formula \eqref{sepapp} when $k\ll1$.}
\label{fig:echosep} 
\end{figure} 

First of all, a direct comparison between the waveforms for $s=2$ (left) and $s=1$ (right) is that, in the case of gravitational perturbations, there appears to be a tiny peak between the ringdown and the arrival of the first echo signal, and such a tiny peak does not appear for the electromagnetic case. In fact, this difference comes from the tiny reflection of the waves happening at the small discontinuity at the core surface in the effective potential (Fig.~\ref{fig:potentials}(a)). This can be understood by observing the arrival time of the tiny peak after the ringdown, which is roughly twice the separation in $x$ between the core discontinuity and the exterior potential peak. However, the amplitudes of such tiny peaks are extremely suppressed by a small $k$, and are not expected to be detectable.

According to the waveforms of the KY model in Figs.~\ref{fig:waveform} and \ref{fig:waveforml3}, we again confirm the presence of a series of echoes following the standard Schwarzschild ringdown phase. In fact, for a sufficiently small $k$, the KY waveforms not only mimic the ringdown phase,{\footnote{During the ringdown phase, our numerical results show that the fundamental QNM frequency deviations relative to the Schwarzschild case, i.e., $\delta\omega\equiv|\omega/\omega_{\textrm{Sch}}-1|$, obtained by fitting the time-domain waveforms via the Prony method are $<10^{-3}$ for the real part, and $<10^{-2}$ for the imaginary part.}} but also follow the Schwarzschild waveform up to some part of the power-law tail. It means that the ringdown phase, including the QNM spectrum, of the KY model is consistent with the Schwarzschild black hole scenario, and it is hard to distinguish between them with black hole spectroscopy.

After the ringdown phase and a portion of the power-law tail, a series of echoes appears in the KY model, with the echo time $\Delta t_{\rm echo}$ increasing as $k$ decreases. Indeed, one can check again that the time separation of successive echoes $\Delta t_{\rm echo}$ is essentially twice the separation between the infinitely high potential barrier and the exterior peak, in units of $x$, i.e., $\Delta t_{\rm echo}\approx 2\Delta x_{\rm sep}$. Here, we estimate $\Delta x_{\rm sep}$ for a small $k$ as{\footnote{The leading order, i.e., the exponential part, is consistent with that of the elapsed coordinate time for light rays introduced in Ref.~\cite{Chen:2024ibc}.}}
\begin{equation}
\Delta x_{\rm sep}/a_0\approx C(\eta)\textrm{exp}\left(\frac{\eta}{4k}\right)\sqrt{k}\,,\label{sepapp}
\end{equation}
where $C(\eta)$ is an $\mathcal{O}(1)$ function independent of $k$. In Fig.~\ref{fig:echosep}, we show how the separation $\Delta x_{\rm sep}$ varies with the parameter $1/k$ for $\eta=1$ (blue) and $\eta=2$ (red). The points are computed numerically from the effective potential \eqref{veff}, and the black curves are obtained from the approximation \eqref{sepapp}. One can see that the numerical results and the approximated formula \eqref{sepapp} perfectly agree when $k\ll1$. Note that the exponential factor in the formula \eqref{sepapp} comes from the redshift of the interior metric \eqref{dense_metric} (see the definition \eqref{k_def} again). Therefore, we can conclude that the time separation between successive echoes is exponentially prolonged due to the strong redshifts inside the KY model.

Using the definition of $k$ \eqref{k_def}, $\sigma \sim n\ell_{\rm Pl}^2$, and $C(\eta)\sim1$, it turns out that the echo emission is strongly redshifted, and the echo time $\Delta t_{\rm echo} =2 \Delta x_{\rm sep}$ can be estimated from \eqref{sepapp} as
\begin{equation}
    \Delta t_{\rm echo} \sim \sqrt{n}\exp \left( \frac{M^2}{nM_{\rm Pl}^2} \right) \times t_{\rm Pl}\,.\label{testimate}
\end{equation}
This timescale is much longer than the cosmological time as long as the mass $M$ of the black hole is much heavier than the Planck mass $M_{\rm Pl}$. For example, in the case of Standard Model with $n\sim 100$, we have $\Delta t \simeq 9$ Gyr for $M \simeq 117
M_{\rm Pl}$ (where $k\approx10^{-3}$), and $\Delta t \simeq \textrm{exp}\left(3.4\times 10^{74}\right)$ Gyr for $M\simeq M_\odot$ (where $k\approx10^{-75}$). Note that in both cases, the condition $k\ll1$ is satisfied. As a result, for the KY model with astrophysical mass scales and a small $k$, the echoes are postponed extremely beyond the Hubble time, making the model consistent with the current null result from echo search.

Note that it takes half of the echo-separation time $\Delta t_{\rm echo}/2 = \Delta x_{\rm sep}$ for waves to enter the surface and reach the central core region, and vice versa. This timescale is also very large, meaning that the detail of the core structure is irrelevant practically in the sense that echoes reflected from there or any emission originating from such a deep quantum regime cannot be detected within a reasonable observational timescale. Similar arguments, but in terms of shadow images, have been made in Ref.~\cite{Chen:2024ibc}, which shows that although light rays propagating inside the KY model can enter the central core region, their propagation timescale with respect to the observer is also exponentially delayed by strong redshifts in the dense region. Therefore, the detailed core structure is irrelevant in the shadow images of the KY model.


\section{Conclusions}\label{sec:conclusion}
Gravitational-wave echoes are expected to be a promising observational signature to probe horizon-scale new physics motivated by quantum gravity. In typical phenomenological models, gravitational-wave echoes are generated by introducing a surface with nonzero reflectivity on top of the Schwarzschild or Kerr spacetime, where the surface is located at a radial distance $L$ away from the horizon radius $a_0$. Then, the echo time is roughly given by $\Delta t_{\rm echo}\approx a_0\ln a_0/L$. In the most extreme scenario in which $L=\ell_{\rm Pl}$, the echo time $\Delta t_{\rm echo}$ would still be of the order of milliseconds for a solar-mass object. This is why the search for such echo signals from the binary black hole mergers has become practically feasible, although so far no strong evidence of such signals has been found \cite{LIGOScientific:2026wpt}. 

However, the non-detection of gravitational-wave echoes does not imply that the object is really a conventional black hole. In fact, even in the absence of an event horizon, if one assumes that the waves can enter and propagate inside the compact object, the echo time will not have the aforementioned logarithmic scaling \cite{Arrechea:2024nlp}. For example, in Ref.~\cite{Zimmerman:2023hua}, by assuming perfect transmission at the star's surface $r=R_s$ and considering isotropic fluid satisfying Tolman-Oppenheimer-Volkoff equation with vanishing pressures at the surface, it was shown that the echo time of a large class of models is characterized by a power-law scaling $\Delta t_{\rm echo}\approx a_0 (L/a_0)^{-p}$, where $L=R_s-a_0$ and $p\ge 1/2$, indicating an echo time within Hubble scale for stellar size objects, but potentially beyond Hubble for intermediate mass ones. In Ref.~\cite{Zimmerman:2023hua}, the authors did not exclude the possibility of a much longer echo time than the power-law scaling in other scenarios. 

In the present paper, we have shown that an echo time much longer than power-law scaling can indeed be realized in the KY model. The exponential factor appearing in the $|g_{tt}|$
component inside the KY model results in an exponentially strong redshift, giving rise to an echo-separation time with exponential scaling: $\textrm{exp}(1/k)\approx\textrm{exp}(a_0^2/\ell_\textrm{Pl}^2)$. Therefore, unless the mass of the KY model is smaller than a hundred times the Planck mass, the echo-separation time would be longer than the Hubble scale. The non-detection of gravitational-wave echoes does not necessarily indicate that the object is a classical black hole.

The KY model is a good candidate for black holes in several aspects. First of all, the model is extremely compact in that the surface is larger than the Schwarzschild radius by just the Planck length scale. In addition to the gravitational-wave echoes whose echo time is beyond Hubble as discussed in this paper, the shadow images produced by it could be hardly distinguishable from the classical black hole images \cite{Chen:2024ibc} due to the strong redshift. Furthermore, the model is consistent with Hawking evaporation \cite{Kawai:2013mda}, and predicts the Bekenstein-Hawking entropy formula \cite{Yokokura:2022phq}. Also, the ratio between viscosity and entropy density of this model is small \cite{Yokokura_viscosity}, which justifies some of our assumptions made in the derivation of the master equation for axial gravitational perturbations.

One then naturally asks whether there are any observational probes that can distinguish the KY model from classical black holes. In Refs.~\cite{Chen:2024ibc}, it was proposed that if the light emission profile around the KY model extends within the shallow region inside the KY surface, there could be extra intensity around the inner shadow of the image, although its magnitude is small for an object with astrophysical scale. 

In terms of gravitational waves emitted from binary mergers, the recently proposed \textit{direct wave} \cite{Oshita:2025qmn,Ma:2026hcb,Su:2026gmp,Nakamoto:2026rah} may be a candidate for such a probe [see also Refs.~\cite{Mino:2008at,Zimmerman:2011dx,Kubota:2026qjh} for relevant studies on the horizon mode and Refs.~\cite{Lu:2025vol,Kankani:2026byb,Chung:2026eph,Dyer:2026yex,Kankani:2026kst,Sun:2026mto} for studies on the possible direct waves in comparable-mass-ratio mergers]. For mergers of classical black holes with, for instance, extreme mass ratio, the direct wave represents the gravitational waves excited by the companion plunging into the black hole near the ergoregion or horizon of the latter. While QNMs predominantly probe the geometry around the light ring, direct waves may carry information about the black hole geometry beyond the light-ring region through the dynamics of the source particle, e.g., frame-dragging in the ergoregion as well as the screening effect of the horizon modes. For the KY model in which no horizon exists, suppose the small companion can enter the shallow region inside the KY surface and remain intact for some time; the direct waves excited from there would carry special features of the model itself. This could provide an opportunity to look for putative KY signatures from the gravitational waveforms near the merger phase in a binary merger event.

Another target that could be a potential probe is associated with the tidal Love numbers, i.e., how much the Love numbers differ from that of a classical black hole, which is precisely zero. Are they exponentially suppressed as well? More importantly, it will be crucial to investigate how such tiny tidal deformations may affect the gravitational waveforms during the inspiral phase. 

In this work, we have studied only the odd-parity mode, which is associated with deformations without changing volume and depends on shear viscosity. Motivated by the small value of the shear viscosity of the KY model \cite{Yokokura_viscosity}, we used the frictionless anisotropic fluid to analyze the odd-parity mode. On the other hand, the even-parity mode occurs with volume changes and depends on bulk viscosity; in addition to the transfer of momentum, it also causes changes in energy density, making it more complex. Note that the bulk viscosity of the KY model is zero \cite{Yokokura_viscosity}. Therefore, it would be interesting to study the even-parity mode by constructing a consistent fluid model, and examine how much the isospectrality (odd mode $=$ even mode in spectra) is violated in the KY model. 

Finally, it would be important to investigate gravitational waves in the KY model based on the semi-classical Einstein equations, rather than on GR coupled with a fluid model. As mentioned in sec. \ref{sec.intro}, this requires considering the time evolution of both the metric and quantum state. As a step toward this quantum analysis, one could examine conformal matter fields and employ the trace component of the semi-classical Einstein equations. Since the trace of the energy-momentum tensor is given by the conformal anomaly, the component is independent of the state. This equation would impose strong constraints on the system of dynamical equations, thereby simplifying the analysis  (see Ref.~\cite{Ho:2023eem} for a static analysis). 

We leave these topics to future work.
\acknowledgments
C.Y.C. is supported by the Special Postdoctoral Researcher (SPDR) Program at RIKEN and RIKEN Incentive Research Grant (Shoreikadai) 2025. 
N.O. was supported by Japan Society for the Promotion of Science KAKENHI Grant No.~JP26K17142.
Y.Y. was partially supported by Japan Society for the Promotion of Science 
(Grants No.21K13929) and by RIKEN iTHEMS Program. 

\appendix
\section{Axial gravitational perturbations with anisotropic fluids}\label{appendixpotential}

In this appendix, we derive the master equation \eqref{waveeq} with the effective potential \eqref{veff} for $s=2$. We consider the perturbations of a static, spherically symmetric spacetime. Without loss of generality, the perturbed spacetime can be described by a non-stationary and axisymmetric (no $\varphi$ dependence) metric as follows \cite{Chandrasekhar:1985kt}
\begin{align}
ds^2=&-e^{2\nu}(dx^0)^2+e^{2\psi}(dx^1-\gamma dx^0-q_2dx^2-q_3dx^3)^2\nonumber\\&+e^{2\mu_2}(dx^2)^2+e^{2\mu_3}(dx^3)^2,\label{metricg}
\end{align}
where $\nu$, $\psi$, $\mu_2$, $\mu_3$, $\gamma$, $q_2$, and $q_3$ are functions of time $t$ ($t=x^0$), radial coordinate $r$ ($r=x^2$) and polar angle $\theta$ ($\theta=x^3$). Because the system is axisymmetric, these functions are independent of the azimuthal angle $\varphi$ ($\varphi=x^1$). In this work, the notation used in Ref.~\cite{Chandrasekhar:1985kt} is strictly followed. The only difference is that the function $\omega$ used in Ref.~\cite{Chandrasekhar:1985kt} is replaced with $\gamma$ here since we have used the notation $\omega$ to denote the frequency of the perturbations. Note that in the background spacetime, which is static and spherically symmetric, $\gamma=q_2=q_3=0$ and functions $\nu$, $\mu_2$, and $\mu_3$ only depend on the radial coordinate $r$, and $e^{2\psi}=e^{2\mu_3}\sin^2\theta$.

We use the tetrad formalism in which a basis associated with the metric \eqref{metricg} can be defined as 
\begin{align}
e^{\mu}_{(0)}&=(e^{-\nu},\quad\gamma e^{-\nu},\quad0,\quad0)\,,\nonumber\\
e^{\mu}_{(1)}&=(0,\quad e^{-\psi},\quad 0,\quad0)\,,\nonumber\\
e^{\mu}_{(2)}&=(0,\quad q_2e^{-\mu_2},\quad e^{-\mu_2},\quad0)\,,\nonumber\\
e^{\mu}_{(3)}&=(0,\quad q_3e^{-\mu_3},\quad 0,\quad e^{-\mu_3})\,,
\end{align}
and
\begin{align}
e_{\mu}^{(0)}&=(e^{\nu},\quad0,\quad0,\quad0)\,,\nonumber\\
e_{\mu}^{(1)}&=(-\gamma e^{\psi},\quad e^{\psi},\quad -q_2e^{\psi},\quad -q_3e^{\psi})\,,\nonumber\\
e_{\mu}^{(2)}&=(0,\quad 0,\quad e^{\mu_2},\quad0)\,,\nonumber\\
e_{\mu}^{(3)}&=(0,\quad 0,\quad 0,\quad e^{\mu_3})\,.
\end{align}
Here, the tetrad indices are enclosed in parentheses to distinguish them from the tensor indices, and the basis satisfies
\begin{align}
e_{\mu}^{(a)}e^{\mu}_{(b)}&=\delta^{(a)}_{(b)}\,,\quad e_{\mu}^{(a)}e^{\nu}_{(a)}=\delta^{\nu}_{\mu}\,,\nonumber\\
e_{\mu}^{(a)}&=g_{\mu\nu}\eta^{(a)(b)}e^{\nu}_{(b)}\,,\nonumber\\
g_{\mu\nu}&=\eta_{(a)(b)}e_{\mu}^{(a)}e_{\nu}^{(b)}\,.
\end{align}
Conceptually, in the tetrad formalism, we project the relevant quantities defined on the coordinate basis of $g_{\mu\nu}$ onto a chosen basis of $\eta_{(a)(b)}$ by constructing the tetrad
basis correspondingly. In practice, $\eta_{(a)(b)}$
is usually assumed to be the Minkowskian matrix
\begin{equation}
\eta_{(a)(b)}=\eta^{(a)(b)}=\textrm{diag}(-1,1,1,1)\,.
\end{equation}
In the tetrad formalism, any vector or tensor field can be projected onto the tetrad frame in which the field is expressed by its tetrad components:
\begin{align}
A_{\mu}&=e_{\mu}^{(a)}A_{(a)}\,,\quad A_{(a)}=e_{(a)}^{\mu}A_{\mu}\,,\nonumber\\
B_{\mu\nu}&=e_{\mu}^{(a)}e_{\nu}^{(b)}B_{(a)(b)}\,,\quad B_{(a)(b)}=e_{(a)}^{\mu}e_{(b)}^{\nu}B_{\mu\nu}\,.
\end{align}

We assume that the geometry is described by the Einstein equation with anisotropic fluids. The energy-momentum tensor reads
\begin{equation}
T_{\mu\nu}=(\rho+p_2)u_\mu u_\nu+(p_1-p_2)x_\mu x_\nu+p_2g_{\mu\nu}\,,\label{anisopf}
\end{equation}
where $\rho$ is the energy density measured by a comoving observer with the fluid velocity $u^\mu$; $p_1$ and $p_2$ are the radial pressure and the tangential pressure, respectively; and $x^\mu$ is the spacelike unit vector orthogonal to $u^\mu$ and angular directions: \begin{equation}
u_\mu u^\mu=-1\,,\qquad x_\mu x^\mu=1\,.\label{fourvelocity}
\end{equation}
At the background level, the four-velocity $u_{(a)}$ and the spacelike unit vector $x_{(a)}$ in the tetrad frame have non-vanishing components only in the $x^0$ and $x^2$ directions, respectively, i.e., $u_{(0)}=x_{(2)}=1$.

Perturbing the energy-momentum tensor \eqref{anisopf} and expressing its components in the tetrad frame, we have
\begin{align}
\delta T_{(a)(b)}=&\,(\rho+p_2)\delta(u_{(a)}u_{(b)})+(\delta\rho+\delta p_2)u_{(a)}u_{(b)}\nonumber\\
&+(p_1-p_2)\delta(x_{(a)}x_{(b)})+(\delta p_1-\delta p_2)x_{(a)}x_{(b)}\nonumber\\&+\delta p_2\eta_{(a)(b)}\,.\label{perturbT}
\end{align}
Furthermore, in the tetrad frame, by linearizing Eq.~\eqref{fourvelocity} and using the orthogonality condition $u_\mu x^\mu=0$, one can obtain
\begin{align}
\delta(u^{(a)} u_{(a)})=0\,,\qquad&\Longrightarrow\qquad \delta u_{(0)}=0\,,\nonumber\\
\delta(x^{(a)} x_{(a)})=0\,,\qquad&\Longrightarrow\qquad \delta x_{(2)}=0\,,\nonumber\\
\delta(u^{(a)} x_{(a)})=0\,,\qquad&\Longrightarrow\qquad \delta u_{(2)}=\delta x_{(0)}\,.
\end{align}

From Eq.~\eqref{perturbT}, one can easily see that $\delta T_{(1)(3)}$ vanishes. According to the Einstein equations, this implies $R_{(1)(3)}=0$. Then, assuming $8\pi G=1$, we have
\begin{align}
R_{(1)(2)}&=\delta T_{(1)(2)}=\left(p_1-p_2\right)\delta x_{(1)}\,,\\
R_{(1)(0)}&=\delta T_{(1)(0)}=\left(\rho+p_2\right)\delta u_{(1)}\,.
\end{align}
The perturbed unit vectors $\delta x_{(1)}$ and $\delta u_{(1)}$ are related to each other as
\begin{align}
\delta T_{(1)(2),r}+\left(\nu_{,r}+2\psi_{,r}+\mu_{3,r}\right)\delta T_{(1)(2)}=e^{\mu_2-\nu}\delta T_{(1)(0),t}\,,\nonumber\label{x1u1relation}\\
\end{align}
where the comma denotes partial derivatives. Eq.~\eqref{x1u1relation} can be obtained by linearizing the conservation equation of the energy-momentum tensor, i.e., $\nabla_\mu \delta T^{\mu\nu}=0$. Therefore, in the general scenario, the axial metric perturbations do not couple to density and pressure perturbations of the fluid, but they couple to fluid velocities $u_{(a)}$. 

In some specific cases, one may have $\delta T_{(1)(2)}=0$, in which the axial gravitational perturbations are not sourced by fluid perturbations. These include
\begin{itemize}
\item The isotropic case where $p_1=p_2$: In the absence of fluid anisotropy, the axial gravitational perturbations completely decouple from the fluid perturbations \cite{Chandrasekhar:1991fi}.
\item The frictionless fluid in which $\delta x_{(1)}=0$ is explicitly imposed: This scenario can be realized by directly assuming that there is no momentum flow in the perturbed stress tensor \cite{Chen:2019iuo,Cardoso:2022whc,Feng:2024ygo}.
\item A specific branch of irrotational fluid: In the absence of fluid vorticity, i.e.,
\begin{equation}
\frac{1}{2}\epsilon^{\alpha\beta\mu\nu}u_{\beta;\mu}u_\nu=0\,,
\end{equation}
one gets $\delta u_{(1)}=0$. Then, from the relation \eqref{x1u1relation}, the solution in the simplest branch corresponds to $\delta x_{(1)}=0$. This scenario was adopted very recently in Refs.~\cite{DOnofrio:2026ulh,Zhao:2026eti,Bhowmik:2026owi}. 
\end{itemize}

Motivated by the small viscosity inside the KY model \cite{Yokokura_viscosity}, we assume frictionless fluids and assume $\delta x_{(1)}=0$, which implies $R_{(1)(2)}=0$. The equations $R_{(1)(2)}=0$ and $R_{(1)(3)}=0$ can be expressed explicitly as
\begin{align}
e^{\nu+\mu_2}\frac{\mathcal{Q}_{,\theta}}{r^4\sin^3\theta}&=-\left(\gamma_{,r}-q_{2,t}\right)_{,t}\,,\label{r12exp}\\
e^{\nu-\mu_2}\frac{\mathcal{Q}_{,r}}{r^2\sin^3\theta}&=\left(\gamma_{,\theta}-q_{3,t}\right)_{,t}\,,\label{r13exp}
\end{align}
respectively, where we have defined
\begin{equation}
\mathcal{Q}\equiv r^2e^{\nu-\mu_2}\left(q_{2,\theta}-q_{3,r}\right)\sin^3\theta\,.
\end{equation} 
Combining Eqs.~\eqref{r12exp} and \eqref{r13exp}, and eliminating the terms containing $\gamma$, one can obtain
\begin{equation}
\left(\frac{e^{\nu-\mu_2}}{r^2}\mathcal{Q}_{,r}\right)_{,r}+\frac{e^{\nu+\mu_2}}{r^4}\left(\frac{\mathcal{Q}_{,\theta}}{\sin^3\theta}\right)_{,\theta}\sin^3\theta=\frac{\mathcal{Q}_{,tt}}{r^2e^{\nu-\mu_2}}\,.\label{qeq}
\end{equation}

To obtain the master equation \eqref{waveeq} with the effective potential \eqref{veff} for $s=2$, we consider the Fourier decomposition and the ansatz
\begin{equation}
\mathcal{Q}(r,\theta)=\mathcal{Q}(r)Y(\theta)\,,
\end{equation}
where $Y(\theta)$ is the Gegenbauer function that satisfies
\begin{equation}
\left(\frac{1}{\sin^3\theta}Y_{,\theta}\right)_{,\theta}=-\left[\ell\left(\ell+1\right)-2\right]\frac{Y}{\sin^3\theta}\,.
\end{equation}
Note that we omit the index $\ell$ in the function $Y(\theta)$. Then, Eq.~\eqref{qeq} can be recast as
\begin{equation}
\left(\frac{e^{\nu-\mu_2}}{r^2}\mathcal{Q}_{,r}\right)_{,r}-\frac{e^{\nu+\mu_2}}{r^4}\left[\ell\left(\ell+1\right)-2\right]\mathcal{Q}=-\omega^2\frac{\mathcal{Q}}{r^2e^{\nu-\mu_2}}\,.
\end{equation}
Finally, we define $\Psi=\mathcal{Q}/r$ and consider the tortoise radius $dx=e^{\mu_2-\nu}dr$. One recovers Eq.~\eqref{waveeq} and the effective potential \eqref{veff} with $s=2$.


\bibliographystyle{mybibstyle}
\bibliography{bib}

\end{document}